# Structured hetero-symmetric quantum droplets

**Yaroslav V. Kartashov,[1,2] Boris A. Malomed,[3,4] and Lluis Torner[1,5]**

[1]ICFO-Institut de Ciències Fotòniques, The Barcelona Institute of Science and Technology, 08860 Castelldefels (Barcelona), Spain

[2]Institute of Spectroscopy, Russian Academy of Sciences, Troitsk, Moscow, 108840, Russia

[3]Department of Physical Electronics, School of Electrical Engineering, Faculty of Engineering, and Centre for Light-Matter Interaction, Tel Aviv University, 69978 Tel Aviv, Israel

[4]Instituto de Alta Investigación, Universidad de Tarapacá, Casilla 7D, Arica, Chile

[5]Universitat Politècnica de Catalunya, 08034, Barcelona, Spain

We predict that Lee-Huang-Yang effect makes it possible to create stable quantum droplets (QDs) in binary Bose-Einstein condensates with a *hetero-symmetric* or *hetero-multipole* structure, i.e., different vorticities or multipolarities in their components. The QDs feature flat-top shapes when either chemical potential $\mu_{1,2}$ of the droplet approaches an edge of a triangular existence domain in the $(\mu_1, \mu_2)$ plane. QDs with different vorticities of their components are stable against azimuthal perturbations, provided that the norm of one component is large. We also present multipole states, in which the interaction with a strong fundamental component balances the repulsion between poles with opposite signs in the other component, leading to the formation of stable bound states. Extended stability domains are obtained for dipole QDs; triple ones exist but are unstable, while quadrupoles are stable in a narrow region. The results uncover the existence of much richer families of stable binary QDs in comparison to states with identical components.

## I. Introduction

Binary Bose-Einstein Condensates (BECs) offer a unique possibility, facilitated by the use of the Feshbach-resonance technique [1-3], to equilibrate intra-component repulsion and inter-component attraction. It was predicted [4,5] that the balance of the weak residual attractive mean-field nonlinearity and additional self-repulsion, induced by the Lee-Huang-Yang (LHY) effect [6], i.e., a correction to the mean-field dynamics produced by Bogoliubov fluctuations, makes it possible to build self-trapped three- and two-dimensional (3D and 2D) states, called *quantum droplets* (QDs), as they are filled by a nearly incompressible quantum liquid. A unique asset of QDs is their stability against the collapse, that destabilizes 2D and 3D solitons of the mean-field type in diverse physical media [7]. The prediction was followed by experimental creation of QDs, with both nearly-2D (oblate) [8,9] and isotropic 3D [10,11] shapes, in mixtures of two different atomic states in $^{39}$K and in an attractive mixture of $^{41}$K and $^{87}$Rb atoms [12]. Still earlier, it was shown that a similar mechanism stabilizes QDs in a single-component gas of dipolar atoms with long-range attraction between them [13-19]. QDs have been also predicted in Bose-Fermi mixtures [20,21], as well as in the form of supersolids [22-24].

The LHY effect strongly depends on the density of states of the Bogoliubov modes, which makes it essentially different in different spatial dimensions [5]. As a result, the corresponding nonlinearity in the modified Gross-Pitaevskii equations (GPEs) is quartic self-repulsion in 3D [4], while in 2D it amounts to a cubic term multiplied by a logarithmic factor [5] (in the 1D limit, it is represented by a quadratic self-attraction term [5]). The logarithmic multiplier implies switching of the 2D nonlinearity from self-attraction to repulsion with the increase of the condensate's density, which makes it possible to create stable states. In addition to the ground-state QDs, interest was drawn to QD vortices. While they are unstable in the dipolar BEC [18], stability areas were found for 3D QDs with vorticities $m = 1$ and $m = 2$ [25]. The LHY-modified GPE in 2D produces stable QDs, at least, up to $m = 5$ [26-28].

To date, studies of QDs with an intrinsic topological structure were focused on binary states with equal vorticities, $m_1 = m_2$ (states with $m_1 = -m_2$ are possible too, but they feature a strongly reduced stability area [27]). Identical components were also assumed in the recently introduced ring-shaped droplet clusters, which may be quasi-stable states too [29,30]. Admitting *hetero-symmetric* states, with $m_1 \neq m_2$ or different symmetries in the components, may essentially expand the variety of modes existing in the system, cf. 1D states with unequal components [31]. In the experiment, different vorticities can be imprinted onto different components of the binary condensate by laser beams which carry the respective vorticities, coupled to different atomic states which represent the two components [32-38]. In nonlinear optics, a similar method made it possible to create two-component vortex solitons with $m_1 \neq m_2$ in photorefractive media [39].

Another possibility to create stable states with unequal vorticities in the two components is to use a trapping potential in cubic nonlinear media [40]. It is relevant too to mention states composed of more than two components with different vorticities [41].

Also predicted [42] and observed [43,44] were binary hetero-symmetric solitons with one fundamental component and another one carrying a dipolar structure, as well as "propellers" with a rotating dipolar structure [45]. Similar to what is mentioned above for binary states with different vorticities, such complexes can be created by means of a pair of laser beams coupling to different atomic states, one beam being structureless, and the other one carrying a dipole or quadrupole transverse profile.

In this work we aim to introduce and explore hetero-symmetric QDs with different vorticities or different multipolarities (fundamental/dipole or fundamental/quadrupole) in the two components. We identify stability domains of such complexes in the plane $(\mu_1, \mu_2)$ of chemical potentials of the components. Quite naturally, the domain shrink with the increase of the vorticity or multipole's order.

The model, based on coupled two-dimensional GPEs with the LHY corrections, is introduced in Section II. Results demonstrating the existence and stability of the hetero-symmetric QDs, obtained by means of numerical and analytical methods, are summarized in Section III. The paper is concluded by Section IV.

## II. The model

As mentioned above, in two dimensions the evolution of the two-component wave function $\psi_{1,2}(x,y,t)$ of a binary BEC is governed by coupled GPEs [5, 26] with the cubic terms multiplied by logarithmic factors that represent the LHY corrections:

$$i\frac{\partial\psi_{1,2}}{\partial t} = -\frac{1}{2}\left(\frac{\partial^2}{\partial x^2} + \frac{\partial^2}{\partial y^2}\right)\psi_{1,2} + (|\psi_{1,2}|^2 - |\psi_{2,1}|^2)\psi_{1,2} + \alpha(|\psi_1|^2 + |\psi_2|^2)\psi_{1,2}\ln(|\psi_1|^2 + |\psi_2|^2). \tag{1}$$

Here the wave functions and coordinates $(x, y)$ are measured in units of $n_0^{1/2}$ and $(n_0|a_s|/a_\perp)^{-1/2}$, respectively, where $n_0$ is the equilibrium density of the 2D QD [5], $a_s$ and $a_\perp$ being the scattering length of atomic collisions and the transverse confinement scale. Here the strength of the difference nonlinearity, $(|\psi_{1,2}|^2 - |\psi_{2,1}|^2)\psi_{1,2}$, is scaled to be 1, while the relative strength $\alpha$ of the LHY terms is kept as a free parameter, which takes generic values $\sim 1$, in the scaled notation. It is found that increase of $\alpha$ leads to practically linear growth of the size of the existence and stability domains, therefore we hereafter fix $\alpha = 1$.

The system conserves total $U$ and partial norms $U_{1,2}$ (which are proportional to numbers of atoms in the two components):

$$U = \iint dx dy(|\psi_1|^2 + |\psi_2|^2) \equiv U_1 + U_2, \tag{2}$$

as well as linear and angular momenta and the Hamiltonian,

$$H = \frac{1}{2}\iint[|\partial\psi_1/\partial x|^2 + |\partial\psi_1/\partial y|^2 + |\partial\psi_2/\partial x|^2 + |\partial\psi_2/\partial y|^2 + (|\psi_1|^2 - |\psi_2|^2)^2 + \alpha(|\psi_1|^2 + |\psi_2|^2)\ln[(|\psi_1|^2 + |\psi_2|^2)/e]]dx dy, \tag{3}$$

in which the last term represents the LHY-modified interaction.

To produce stationary droplet solutions of Eq. (1), we used the Newton iterative method. The evolution of QDs in time was simulated by means of split-step fast Fourier method applied to Eq. (1).

## III. Results

First, we address axisymmetric QDs with different vorticities $m_{1,2}$ and chemical potentials $\mu_{1,2} < 0$ of their components. In polar coordinates $(r, \varphi)$, the corresponding solutions are sought for as

$$\psi_{1,2} = w_{1,2}(r)e^{im_{1,2}\varphi - i\mu_{1,2}t},$$

where $w_{1,2}(r)$ are real radial profiles of the wave function. The stability of the stationary solutions is analysed below by considering perturbed solutions,

$$\psi_{1,2} = (w_{1,2} + u_{1,2}e^{\delta t + in\varphi} + v_{1,2}^*e^{\delta^* t - in\varphi})e^{-i\mu_{1,2}t + im\varphi} \tag{4}$$

with azimuthal index $n$ and growth rate $\delta$. The substitution of this in Eq. (1) and derivation of the corresponding Bogoliubov – de Gennes equations, by linearization of the GPEs for small perturbations $\sim u_{1,2}, v_{1,2}$, leads to an eigenvalue problem for the growth rate:

$$i\delta u_{1,2} = -(1/2)[u''_{1,2} + r^{-1}u'_{1,2} - (m_{1,2} + n)^2 r^{-2}u_{1,2}] - \mu_{1,2}u_{1,2} + w^2_{1,2}(2u_{1,2} + v_{1,2}) - w^2_{2,1}u_{1,2} - w_1w_2(u_{2,1} + v_{2,1}) + \alpha u_{1,2}(w_1^2 + w_2^2)\mathcal{L} + \alpha w_{1,2}(1+\mathcal{L})[w_1(u_1 + v_1) + w_2(u_2 + v_2)],$$

$$i\delta v_{1,2} = +(1/2)[v''_{1,2} + r^{-1}v'_{1,2} - (m_{1,2} - n)^2 r^{-2}v_{1,2}] + \mu_{1,2}v_{1,2} - w^2_{1,2}(2v_{1,2} + u_{1,2}) + w^2_{2,1}v_{1,2} + w_1w_2(u_{2,1} + v_{2,1}) - \alpha v_{1,2}(w_1^2 + w_2^2)\mathcal{L} - \alpha w_{1,2}(1+\mathcal{L})[w_1(u_1 + v_1) + w_2(u_2 + v_2)], \tag{5}$$

with $\mathcal{L} \equiv \ln(w_1^2 + w_2^2)$.

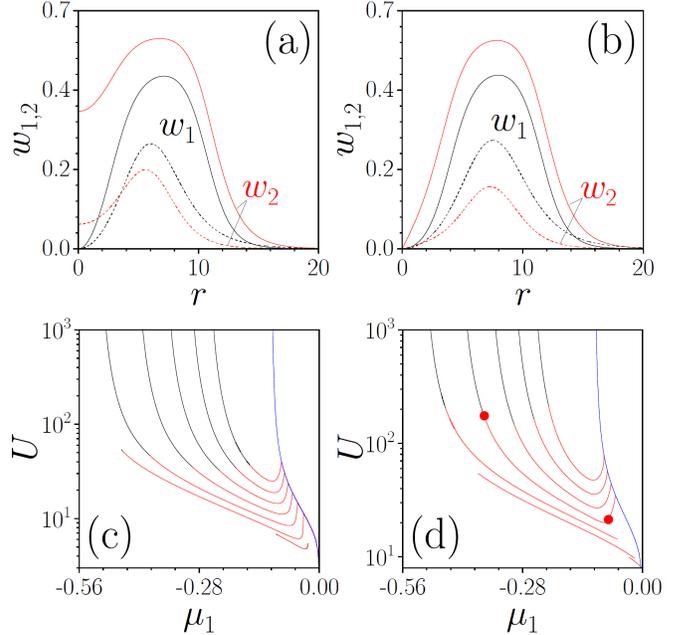

Fig. 1. Examples of axisymmetric QDs with vorticities $(m_1, m_2) = (2, 0)$, i.e., *semi-vortices*. (a) and (2,1) (b) in components with $\mu_{1,2} = (-0.08, -0.02)$ (dashed lines, unstable states) and $\mu_{1,2} = (-0.37, -0.20)$ (solid lines, stable states). The total norm of QDs vs. $\mu_1$ at fixed $\mu_2 = -0.08, -0.1, -0.15, -0.2, -0.25, -0.3$, and $-0.35$, from the bottom to top, is shown for QDs with $(m_1, m_2) = (1,0)$ and $(m_1, m_2) = (2,0)$ in (c) and (d), respectively. Blue solid lines are $U(\mu_1)$ curves for the single-component states with $w_1 \neq 0$, $w_2 = 0$ and $m_1 = 1$ (c) or $m_1 = 2$ (d). Dots in (d) represent QDs shown in (a).

Representative profiles of QDs with $(m_1, m_2) = (2, 0)$ and $(2, 1)$ are shown in Figs. 1(a,b) (the complex of the former type, with vorticity carried by a single component, may be naturally called a *half-vortex* [46] or *semi-vortex* [47]). The increase of either $|\mu_1|$ or $|\mu_2|$ typically leads to growth of amplitudes of both components, which feature broad shapes (solid lines in the plots), due to the effect of the LHY-induced logarithmic factor in the nonlinearity in Eq. (1). Even if one component has zero vorticity, it develops a ring shape with a pronounced minimum at the centre. Decrease of $|\mu_{1,2}|$ at fixed $\mu_{2,1}$ leads to gradual vanishing of $\psi_{2,1}$. However, the limit case in which one component vanishes, is not adequately modelled by Eq. (1) also [31].

In contrast to the case of QDs with identical components [5,27], in our case total norm $U$ is a non-monotonous function of the chemical potentials, as shown in Figs. 1(c,d) for $U(\mu_1)$ which pertains to $(m_1, m_2) = (1,0)$ and $(2,0)$. For sufficiently large $|\mu_2|$, the total norm diverges when the QDs get broad as $|\mu_1|$ increases. In contrast, when $|\mu_1|$ decreases, $U(\mu_1)$ merges into the blue line representing a scalar solution of Eq. (1) with $\psi_1 \neq 0$, $\psi_2 \equiv 0$. Stable branches [i.e., parts of $U(\mu_1)$ dependence corresponding to dynamically stable states shown by black colour in Figs. 1(c,d)] start above a certain critical norm. Similar results were obtained for other sets of $m_{1,2}$.

Existence domains for QDs and the results of the corresponding linear-stability analysis, based on numerical solution of Eq. (5), are summarized in Fig. 2(a-c). Vortex QDs exist in triangular regions in the $(\mu_2, \mu_1)$ plane. Close to the top and right edges of the triangles, $\psi_2$ and $\psi_1$ vanish, respectively, for any combination of $m_1$ and $m_2$. At the left edge of the triangular regions, both components acquire flat-top shapes, with different amplitudes for $\mu_1 \neq \mu_2$, hence one expects a stability region close to this border of the existence domain.

In addition to the numerical results, the left edge of the triangular existence area, identical for all combinations of $m_{1,2}$, can be found in an exact analytical form. To this end, in the flat-top regime, with derivatives of $\psi_{1,2}$ neglected in the lowest approximation, Eq. (1) is combined with the condition of the conservation of the formal Hamiltonian, $h$, in the stationary version of Eq. (1) in the same flat-top limit [ $h$ is given by the last two terms in the Hamiltonian density corresponding to Eq. (3)], cf. Ref. [25]. A straightforward calculation, based on the latter principle, yields an exact result, *viz.*, a parametric form of the left edge of the triangle, which completely coincides with its numerically found counterpart displayed in Fig. 2:

$$\mu_{1,2} = \pm \alpha^{1/2} n [\ln(1/e^{1/2} n)]^{1/2} + \alpha n \ln n, \qquad (6)$$

where total density $n \equiv w_1^2 + w_2^2$ varies in the range of $n_{\min} = e^{-(1/2 + 1/\alpha)} < n < n_{\max} = e^{-1/2}$. Positions of the left and bottom vertices of the triangle, at which the density of either component vanishes, can be obtained from Eq. (6). In particular, the coordinates of the left vertex are $\mu_1 = -(\alpha/2) e^{-1/2 - 1/\alpha}$ and $\mu_2(\alpha = 1) \approx -0.612$ (the latter value is a numerical solution of an algebraic equation). While in Fig. 2(a,c) it is possible to produce numerical data close to the triangle's vertex at $\mu_{1,2} = 0$, in Fig. 2(b) this is challenging because the existence domain strongly shrinks.

By and large, vortex QDs are prone to azimuthal instabilities with low azimuthal perturbation indices $n$, which, however, may be suppressed by the LHY effect [26-28]. Lines of different colours in Fig. 2(a-c) indicate borders at which the instabilities with different values of the perturbation azimuthal index $n$ in Eq. (4) disappear ($n = 0$ does not produce instability). Dependencies of the real part of the perturbation growth rates, $\delta_{re}$, on $\mu_1$ for fixed $n$ are presented in Fig. 3, using the same colour coding for $n$ as in Fig. 2. These results predict a possibility to observe different dynamical scenarios by applying specific perturbation modes, corresponding to the respective values of $n$, to unstable QDs. In addition to the prediction of the exact border of the stability domain, the dependencies $\delta_{re}(\mu_1)$ allow us to clearly identify the azimuthal index of the fastest growing perturbation mode. A final conclusion is that QDs are stable in crescent-shaped domains $s$ in Fig. 2, where $\delta_{re} = 0$ for all $n$.

We thus find that QDs with larger vorticities are vulnerable to a broader set of azimuthal perturbations. While the mode with $(m_1, m_2) = (1, 0)$ is subject to instabilities with $1 \leq n \leq 3$, its counterpart with $(m_1, m_2) = (2, 1)$ is destabilized by $1 \leq n \leq 5$. The most destructive perturbation for the QD modes shown here has $n = 2$. Nevertheless, for QDs of all the types considered here a stability domain $s$ is identified in Figs. 2(a-c), getting narrower for higher vorticities, see Figs. 2(a) and (c). Note that the stability domains cover a variety of QDs with all values of shares of the components' norms, $U_{1,2}/U$, as they extend from the left vertex (with $w_2 \to 0$) to the bottom one (with $w_1 \to 0$).

This result substantially expands the class of stable 2D QDs found in previous works. We have found similar but narrower stability domains for QDs with vorticities $m_1, m_2$ up to 5 and $m_1 \neq m_2$. Generally, the possibility of having narrow but nonvanishing stability areas for vortex solitons with $m > 1$ is a known feature of 2D models with completing nonlinearities [48]. We have also found that QDs with component vorticities of opposite signs, such as $(m_1, m_2) = (2, -1)$, are always unstable.

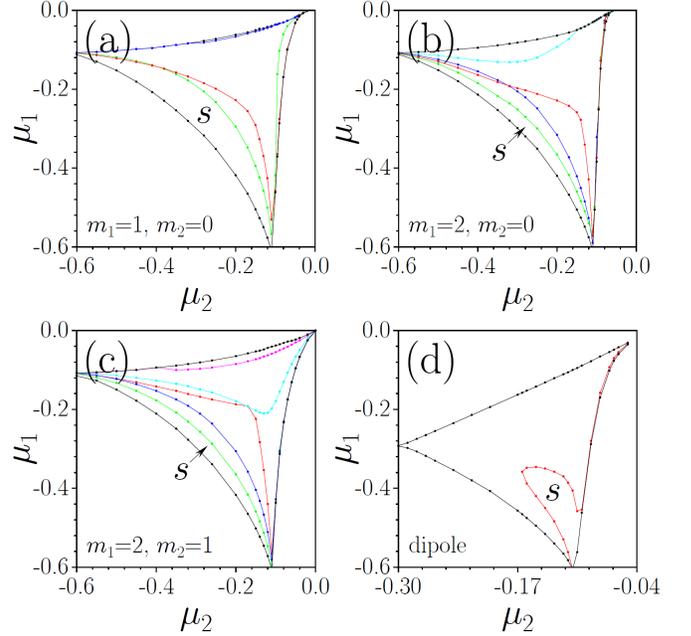

Figure 2. Stability and existence domains for QDs with different vorticities in the two components (a)-(c) and (semi-) dipole QDs (d) in the $(\mu_2, \mu_1)$ plane. QDs exist in triangular regions bounded by lines with black dots [the analytical form of the left edge is given by Eq. (6)]. Color lines with dots in (a)-(c) are borders of domains where the instability with azimuthal indices $n$ for vortex QDs disappear upon the increase of $|\mu_1|$ ($n = 1$ - red, $n = 2$ - green, $n = 3$ - blue, $n = 4$ - cyan, $n = 5$ - magenta). QDs are fully stable in domains marked by $s$, located close to the left edge of the triangle. For dipoles, the stability domain $s$ in (d) is bounded by the red line with dots.

Stable and unstable evolution of QDs, produced by direct simulations of Eq. (1), is displayed in Fig. 4. Unstable modes break into sets of fundamental (zero-vorticity) QDs, whose components typically carry different norms and fly away in tangential directions, to conserve the angular momentum. The breakup, induced by different perturbation eigenmodes, is shown in Figs. 4(a-d). Further, examples of the stable evolution of QDs with $(m_1, m_2) = (2, 0)$ (semi-vortex) and $(m_1, m_2) = (2, 1)$ are displayed in Figs. 4(e,f). Even in the presence of considerable perturbations, such QDs keep their original shape over indefinitely long times.

Another basic finding is that stable hetero-symmetric QDs can be built with different multipolarities, rather than vorticities, in the two components. In such complexes one component is a set of several poles with opposite signs (local QDs), which do not escape under the action of mutual repulsion, as they are nonlinearly coupled to the other (structureless) component, whose wave function does not feature a sign-changing pattern. The simplest representative of this class of composite QDs is the (semi-) dipole shown in Fig. 5(a). It is produced by Eq. (1) in the form of $\psi_{1,2} = w_{1,2}(x, y) e^{-i\mu_{1,2} t}$ with real functions $w_{1,2}$. The corresponding initial guess for the dipole component was taken as $\psi_1 \sim x\exp(-\beta r^2)$, $\beta > 0$. Dependences of the norm on chemical potentials for such QDs are qualitatively similar to those for the vortices, see Figs. 1(c,d). The existence domain for the semi-dipole QDs is shown in Fig. 2(d). The domain is substantially narrower than its counterpart for the axisymmetric vortex droplets. The dipole component of this composite mode vanishes at the right edge of the triangular existence domain, while the structureless component vanishes at the upper edge. The left edge corresponds to the structure in which both components develop broad shapes. In the latter case, the distance between poles of the dipole component, $\psi_1$, substantially increases, while component $\psi_2$ transforms into two weakly overlapping in-phase fragments. Therefore, such QDs may be considered as a bound state of two fundamental QDs, whose $\psi_1$ and $\psi_2$ components are juxtaposed so as to be out-of-phase and in-phase, respectively.

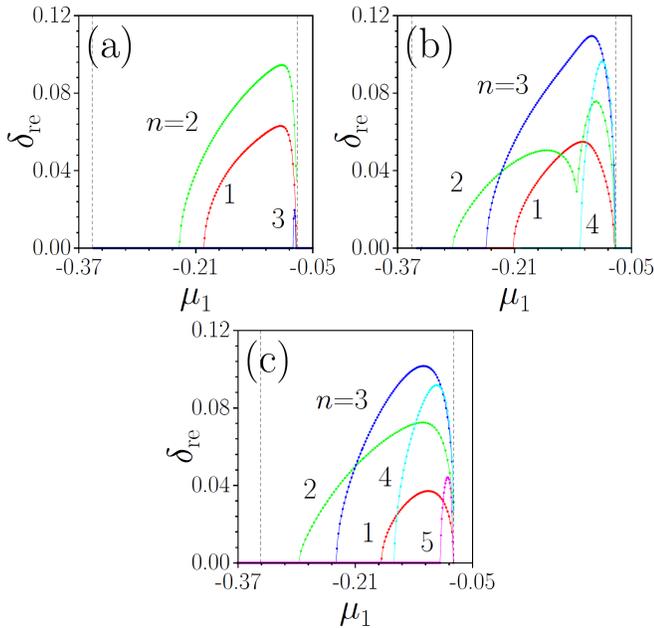

Fig. 3. Instability growth rates, i.e., real parts of eigenvalues $\delta$, for several fixed values of the perturbation azimuthal index $n$, vs. $\mu_1$ for QDs with $(m_1, m_2) = (1, 0)$ (a), $(2, 0)$ (b), and $(2, 1)$ (c). In (a,b) $\mu_2 = -0.25$, and in (c) $\mu_2 = -0.26$. Vertical dashed lines denote borders of the existence domain. The QDs are stable in regions where $\delta_{\mathrm{re}} = 0$ for all values of $n$.

Dipole QDs are stable in a limited part of their existence domain with a complex shape, labelled $s$ in Fig. 2(d), which is bounded by the red line. It is close to the bottom vertex of the existence triangle, in contrast to the stability domains for the vortices displayed above. The cause of the instability of the broad modes is elongation of the nodal line between the poles and emergence of a pattern resembling a quasi-1D dark soliton. When the nodal line becomes long enough, a snake instability sets in (which is typical for such states [49,50]), accompanied by oscillations of amplitudes of the two poles of the dipole. An example of such instability, which is actually rather weak, is presented in Fig. 5(b), while the evolution of a stable semi-dipole is shown in Fig. 5(a). Note also that an additional, very narrow, stability domain for semi-dipoles is stretched along the entire right edge of the existence triangle in Fig. 2(d). In the latter case, a weak dipole component is stabilized by the stronger structureless one.

We have also found QDs with a more complex structure. These are unstable *tripoles*, with one component built of three poles with alternating signs, set along a line, and (semi-) quadrupoles, with the quadrupole structure carried by a single component, see an example in Fig. 5(c). The existence domain for the semi-quadrupoles practically coincides with that for dipoles [see Fig. 2(d) above], while the stability takes place in a narrow area adjacent to the right edge of the existence triangle. The evolution of a stable semi-quadrupole is shown in Fig. 5(c), while Fig. 5(d) illustrates the decay of an unstable solution, in which four separate poles fuse into a broad pattern.

## IV. Conclusion

In this work, we have uncovered a ramified variety of structured hetero-symmetric QDs in binary BECs, with different vorticities or multipolarities of the two components. In particular, these states include semi-vortices and semi-multipoles, with the respective structure carried by a single component, as well as hetero-vortical complexes, with unequal nonzero vorticities in both components. The stability of these states against azimuthal and symmetry-breaking perturbations is an example of novel phenomena offered by the LHY (Lee-Hung-Yang) corrections to the mean-field dynamics. In two-dimensional settings, the LHY effect adds the logarithmic factor to the cubic self- and cross-interactions. Undoing rescaling that led to Eq. (1), an estimate for the number of atoms in the QD with relevant physical parameters is $\sim 10^4 - 10^5$, cf. Refs. [26,27].

Challenging extensions of this work may be the exploration of three-dimensional composite modes, including more complex topologically organized ones, such as skyrmions [51-55]. Another challenging extension may be investigation of the hetero-symmetric and hetero-multipole modes, both two- and three-dimensional ones, in the framework of the full multi-body quantum theory, rather than reducing the quantum effects to the LHY corrections, cf. Ref. [56]. We also anticipate the relevance of the application of the hetero-symmetry concept to complexes hold in trapping potentials, cf. Refs. [46,57].

**Acknowledgments**

Y.V.K. and L.T. acknowledge support from the Government of Spain (Severo Ochoa CEX2019-000910-S), Fundació Cellex, Fundació Mir-Puig, Generalitat de Catalunya (CERCA). Y.V.K. acknowledges partial support of this work by program 1.4 of Presidium of the Russian Academy of Sciences, "Topical problems of low temperature physics." The work of B.A.M. is supported, in part, by the Israel Science Foundation through grant No. 1286/17, and by grant No. 2015616 from the joint program of NSF and BSF (Binational (US-Israel) Science Foundation).

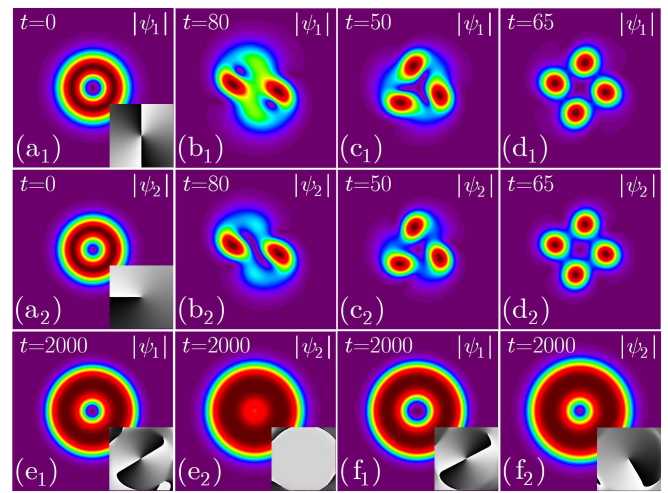

Fig. 4. (a-d) The evolution of an unstable QD with $(m_1, m_2) = (2, 1)$ and $\mu_{1,2} = (-0.118, -0.200)$ is illustrated by profiles $|\psi_{1,2}|$ taken at different times, with insets depicting the phase of the wavefunction. The instability is induced by perturbations with azimuthal indices [see Eq. (4)] $n = 2$ (b), $n = 3$ (c), and $n = 4$ (d). Examples of stable QDs with $(m_1, m_2) = (2, 0)$ (a semi-vortex) in (e) and $(2, 1)$ in (f), surviving long evolution at $\mu_{1,2} = (-0.375, -0.200)$. Here and in Fig. 5, all patterns are displayed in domain $0 < |x|, |y| < 22$.

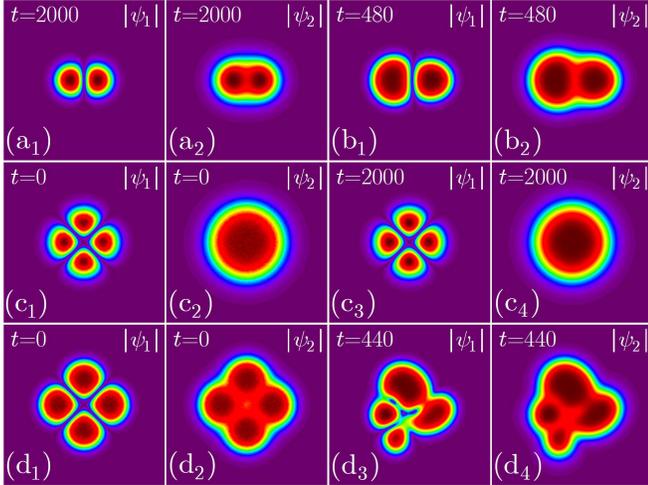

Fig. 5. (a) Stable evolution of a semi-dipole with $\mu_{1,2} = -(0.39, 0.14)$. (b) An unstable semi-dipole with $\mu_{1,2} = (-0.448, -0.15)$. (c) Stable evolution of a semi-quadrupole, with $\mu_{1,2} = (-0.366, -0.1)$. (d) Evolution of an unstable semi-quadrupole with $\mu_{1,2} = (-0.4, -0.17)$.